\documentclass[twocolumn, table]{aastex63}
\usepackage{xcolor}
\usepackage[utf8]{inputenc}
\usepackage{comment}
\usepackage{gensymb}

\shorttitle{Mapping GRBs}
\shortauthors{Espinoza, S. et al.}
\graphicspath{{./}{figures/}}

\begin{document}

\title{Mapping Gamma-Ray Bursts: Distinguishing Progenitor Systems Through Machine Learning}

\author[0009-0003-5478-2378]{Sharleen N. Espinoza}
\affiliation{CCS-2, Computational Physics and Methods, Los Alamos National Lab, Los Alamos, NM 87544}
\author[0000-0003-1707-7998]{Nicole M. Lloyd-Ronning}
\affiliation{CCS-2, Computational Physics and Methods, Los Alamos National Lab, Los Alamos, NM 87544}
\author[0000-0002-6548-5622]{Michela Negro}
\email{michelanegro@lsu.edu}
\affiliation{Department of Physics \& Astronomy, Louisiana State University, Baton Rouge, LA 70803, USA}
\author[0000-0002-4854-8636]{Roseanne M.~Cheng}
\affiliation{CCS-2, Computational Physics and Methods, Los Alamos National Lab, Los Alamos, NM 87544}
\affiliation{Department of Physics and Astronomy, University of New Mexico, Albuquerque, NM 87106, USA}
\author[0000-0003-3842-4493]{Nicol\'o Cibrario}
\affiliation{Istituto Nazionale di Fisica Nucleare, Sezione di Torino, Via Pietro Giuria 1, 10125 Torino, Italy}%
\affiliation{Dipartimento di Fisica, Università degli Studi di Torino, Via Pietro Giuria 1, 10125 Torino, Italy}




\begin{abstract}
We present an analysis of gamma-ray burst (GRB) progenitor classification, through their positions on a Uniform Manifold Approximation and Projection (UMAP) plot, constructed by \citet{negro2024}, from Fermi-GBM waterfall plots.  The embedding plot has a head-tail morphology, in which GRBs with confirmed progenitors (e.g. collapsars vs. binary neutron star mergers) fall in distinct regions.  We investigate the positions of various proposed sub-populations of GRBs, including those with and without radio afterglow emission, those with the lowest intrinsic luminosity, and those with the longest lasting prompt gamma-ray duration. The radio-bright and radio-dark GRBs fall in the head region of the embedding plot with no distinctive clustering, although the sample size is small. Our low luminosity GRBs fall in the head/collapsar region.  A continuous duration gradient reveals an interesting cluster of the longest GRBs ($T_{90} > 100s$) in a distinct region of the plot, possibly warranting further investigation. 
\end{abstract}



\section{Introduction}
\label{sec:intro}
Gamma-ray bursts (GRBs) are the most luminous explosions in the universe, powered by progenitor systems involving the collapse of a massive star \citep[collapsar, e.g.][]{MW99} or a compact binary merger \citep{Berg14}. Short-duration GRBs (with prompt gamma-ray emission $\lesssim 2s$) are typically associated with binary neutron star (BNS) mergers, while long-duration GRBs (prompt gamma-ray duration $\gtrsim 2s$) are linked to collapsars. 
Although both produce relativistic jets and prompt gamma-ray emission, they differ not only in duration, but also in the hardness of their spectrum, total isotropic energy release, and their afterglow properties. Hence, although the duration classification is a useful guideline, there remain many open questions about the nature of, and how best to determine, a GRB progenitor. 

  For example, \cite{lloydronning2017,lloydronning2019, chakraborty2022} showed that GRBs with radio afterglows (radio-bright) have longer prompt gamma-ray durations and higher isotropic energies than those without this emission (radio-dark), and suggested that these sub-classes may originate from distinct progenitor systems and/or environments \citep{lloydronning2022}. Low-luminosity GRBs (\textit{ll}GRBs), meanwhile, are typically smooth single-episode bursts believed to result from failed jets or mildly relativistic shock breakouts in collapsars \citep{liang07, bromberg2011, nakar2015}. These observational distinctions may also trace back to fundamental differences in the underlying progenitor systems.  \\

Recently, in an effort to more carefully classify GRB progenitors, \citet{negro2024} constructed a series of GRB waterfall plots, which incorporate detailed light curve variability and time resolved spectral information.  They built a convolutional autoencoder taking these multidimentional waterfall plots as input and reducing the total dimensionality to 30 and, subsequently, to two- and three-dimensions,
by means of the Uniform Manifold Approximations and Projections \citep[UMAPs, e.g.][]{osti_10104557}. This projection reveals a distinct head-tail structure that captures the classic hard-short/long-soft GRB separatin \citep{Kouv93}, but in a more nuanced manner that can potentially better distinguish between the progenitors of GRB populations.
In this study, we examine the positions of different subclasses of GRBs on the UMAP embedding plot, including radio bright and dark GRBs, \textit{ll}GRBs, and those with the longest prompt gamma-ray durations.

\section{Methods} 
\label{sec:methods}
As described in the introduction, we utilize the two-dimensional UMAP embedding constructed by \citet{negro2024}, derived from Fermi-GBM Continuous Time-Tagged Event (CTTE) data (although the results below are valid in three dimensions as well; see \url{https://nmik.github.io/SmartWaterfalls/plotly/grbs_mv10umap_plotly_v0.html}).
The UMAP projection reveals a prominent head-tail morphology, where those GRBs observationally associated with supernovae fall in one region of the plot, while the shortest duration GRBs, including the one observationally associated with a binary neutron star merger (GRB170817), fall in another region.

To investigate how physically motivated sub-classes align with the overall embedding plot, we overlaid three subsets:
\begin{enumerate}
    \item Thirteen GRBs with radio follow-up from \citet{lloydronning2017},
    \item Five \textit{ll}GRBs ($L < 10^{49} erg s^{-1}$) from \citet{dong2025}, and
    \item An overall duration gradient using Fermi-GBM $T_{90}$ values for the full embedded sample ($N=2511$).
\end{enumerate}


\section{Results}
\label{sec:results}
\begin{figure*}
\includegraphics[width=\linewidth]{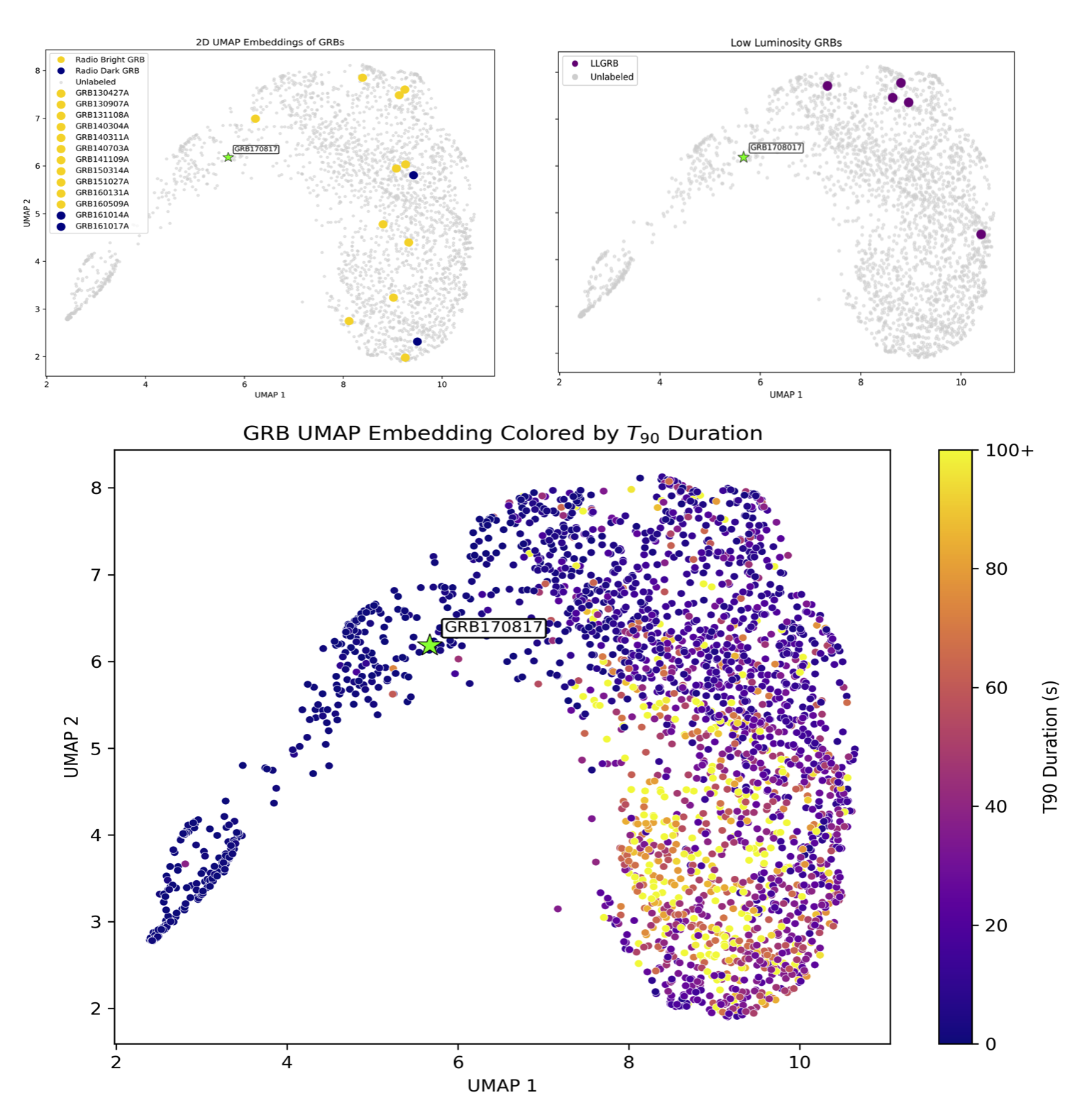}

    \caption{(a) {\bf Top left:} GRBs with known radio afterglow classification; radio-bright GRBs are shown in gold, radio-dark in navy. (b) {\bf Top right:} Low-luminosity GRBs (\textit{ll}GRBs) with $L < 10^{49}$ erg s$^{-1}$, overlaid in purple (c) {\bf Bottom: }Continuous gradient of $T_{90}$ durations for all 2511 GRBs; the longest duration GRBs appear to cluster in a distinct region in the head.}
    \label{fig:RNAAS_fig}
\end{figure*}

Figure \ref{fig:RNAAS_fig} displays our results. 
The 13 long GRBs with known radio afterglow classification are shown in the top left panel. As expected, nearly all fall within or near the "head" region of the embedding, associated with collapsar-like progenitors. 
No clear spatial separation is observed between radio-bright and radio-dark events.  One radio bright data point falls closer to the ``tail'' region of the embedding plot, and may be worth a more detailed examination (particularly in light of the suggestion that these radio bright GRBs are massive stars in interacting binary systems \citep{lloydronning2022}). 

The top right panel shows the location of 5 \textit{ll}GRBs ($L<10^{49} erg \ s^{-1}$, four of which occupy a peripheral region of the head \citep[this region is also occupied by a few other noteworthy GRBs, including two long GRBs with observed kilonovae emission; see Figure 1 of][]{negro2024}.

The bottom panel overlays a continuous color map of $T_{90}$ values for all 2511 GRBs in the dataset. A clear duration gradient spans the head-tail axis, with short bursts concentrated in the tail and longer bursts clustering in the head. The longest duration GRBs ($T_{90} \gtrsim 100s$) appear tightly grouped within a subregion of the head, suggesting a potentially distinct population.

\section{Conclusion}
\label{sec:conclusion}
We have examined whether different sub-classes of GRBs show meaningful structure on the UMAP embedding plot from \citet{negro2024}. Thirteen long duration GRBs with radio afterglows were overlaid onto the embedding, all clustering in or near the head region, but with no definitively clear separation between radio-bright and radio-dark events. Similarly, five \textit{ll}GRBs were located near the head or ``transition'' zone between the head and tail. However, the small sample sizes for both groups limit the statistical power of these comparisons and preclude any firm conclusions about distinct progenitor pathways.

Interestingly, we did find that the longest duration GRBs with $T_{90} \gtrsim 100s$ seem to cluster in a particular region of the head of the plot.  In future work, we will explore in more detail the nature of this clustering, as well as any trends related to the three distinct time resolutions of the waterfall plots.   We emphasize again the usefulness of the spectral and timing information encoded into the embedding plot, potentially helping us uncover the nature of GRB progenitors.

\section{Acknowledgments}
\label{sec:ack}


We thank Eric Burns, whose contributions to the ML part of this work were invaluable.
We also thank Omer Bromberg for insightful conversations about GRB progenitors.  
S.N.E. and N.L.R. acknowledge support from Carlos Di Stefano and LANL's LDRD program, project numbers 20230115ER and 20230217ER.   Los Alamos National Laboratory is operated by Triad National Security, LLC, for the National Nuclear Security Administration of U.S. Department of Energy (Contract No. 89233218CNA000001). LA-UR-25-28441


\bibliographystyle{aasjournal}
\bibliography{bib}



\setlength{\arrayrulewidth}{0.5mm}
\setlength{\tabcolsep}{3pt}
\renewcommand{\arraystretch}{1.3}

\end{document}